%% file: main.tex
\documentclass[a4paper,11pt]{article}
\usepackage{jinstpub} 
\usepackage{lineno}
\usepackage{subcaption}
\usepackage{xcolor}

\title{Real-time Signal Detection for Cyclotron Radiation Emission Spectroscopy Measurements using Antenna Arrays}

\collaboration{\includegraphics[height=17mm]{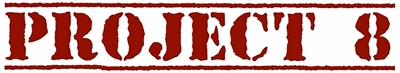}\\[6pt]
The Project 8 collaboration}

\include{author.tex}

\emailAdd{ziegler@psu.edu}

\abstract{
Cyclotron Radiation Emission Spectroscopy (CRES) is a technique for precision measurement of the energies of charged particles, which is being developed by the Project 8 Collaboration to measure the neutrino mass using tritium beta-decay spectroscopy. Project 8 seeks to use the CRES technique to measure the neutrino mass with a sensitivity of 40~meV, requiring a large supply of tritium atoms stored in a multi-cubic meter detector volume. Antenna arrays are one potential technology compatible with an experiment of this scale, but the capability of an antenna-based CRES experiment to measure the neutrino mass depends on the efficiency of the signal detection algorithms. In this paper, we develop efficiency models for three signal detection algorithms and compare them using simulations from a prototype antenna-based CRES experiment as a case-study. The algorithms include a power threshold, a matched filter template bank, and a neural network based machine learning approach, which are analyzed in terms of their average detection efficiency and relative computational cost. It is found that significant improvements in detection efficiency and, therefore, neutrino mass sensitivity are achievable, with only a moderate increase in computation cost, by utilizing either the matched filter or machine learning approach in place of a power threshold, which is the baseline signal detection algorithm used in previous CRES experiments by Project 8.
}


\begin{document}
\maketitle
\flushbottom
\input{sections/1-introduction.tex}
\input{sections/2-a_realtime_detection_routine}

\input{sections/3-signal-classifiers}
\input{sections/4-methods.tex}

\input{sections/5-detection_performance_results.tex}

\input{sections/6-conclusion.tex}

\input{sections/appendix.tex}

\acknowledgments

This material is based upon work supported by the following sources: the U.S. Department of Energy Office of Science, Office of Nuclear Physics, under Award No.~DE-SC0020433 to Case Western Reserve University (CWRU), under Award No.~DE-SC0011091 to the Massachusetts Institute of Technology (MIT), under Field Work Proposal Number 73006 at the Pacific Northwest National Laboratory (PNNL), a multiprogram national laboratory operated by Battelle for the U.S. Department of Energy under Contract No.~DE-AC05-76RL01830, under Early Career Award No.~DE-SC0019088 to Pennsylvania State University, under Award No.~DE-FG02-97ER41020 to the University of Washington, and under Award No.~DE-SC0012654 to Yale University; the National Science Foundation under Award No.~PHY-2209530 to Indiana University, and under Award No.~PHY-2110569 to MIT; the Cluster of Excellence "Precision Physics, Fundamental Interactions, and Structure of Matter" (PRISMA+ EXC 2118/1) funded by the German Research Foundation (DFG) within the German Excellence Strategy (Project ID 39083149); the Karlsruhe Institute of Technology (KIT) Center Elementary Particle and Astroparticle Physics (KCETA); Laboratory Directed Research and Development (LDRD) 18-ERD-028 and 20-LW-056 at Lawrence Livermore National Laboratory (LLNL), prepared by LLNL under Contract DE-AC52-07NA27344, LLNL-JRNL-840701; the LDRD Program at PNNL; and Yale University.  A portion of the research was performed using the HPC cluster at the Yale Center for Research Computing. Additionally, a portion of the computations for this research were performed on the Pennsylvania State University’s Institute for Computational and Data Sciences’ Roar supercomputer, and was supported by the National Science Foundation Award No.~PHY-2018280 to Pennsylvania State University.

\bibliographystyle{JHEP}
\bibliography{biblio.bib}
\end{document}

%% file: author.tex
\newcommand{\mitt}{Laboratory for Nuclear Science, Massachusetts Institute of Technology, Cambridge, MA 02139, USA}
\newcommand{\mainz}{Institute for Physics, Johannes Gutenberg University Mainz, 55128 Mainz, Germany}
\newcommand{\psu}{Department of Physics, Pennsylvania State University, University Park, PA 16802, USA}
\newcommand{\cwru}{Department of Physics, Case Western Reserve University, Cleveland, OH 44106, USA}
\newcommand{\uw}{Center for Experimental Nuclear Physics and Astrophysics and Department of Physics, University of Washington, Seattle, WA 98195, USA}
\newcommand{\pnnl}{Pacific Northwest National Laboratory, Richland, WA 99354, USA}
\newcommand{\yale}{Wright Laboratory, Department of Physics, Yale University, New Haven, CT 06520, USA}
\newcommand{\llnl}{Lawrence Livermore National Laboratory, Livermore, CA 94550, USA}
\newcommand{\iu}{Center for Exploration of Energy and Matter and Department of Physics, Indiana University, Bloomington, IN 47405, USA}
\newcommand{\kit}{Institute for Astroparticle Physics, Karlsruhe Institute of Technology, 76021 Karlsruhe, Germany}

\newcommand{\uta}{Department of Physics, University of Texas at Arlington, Arlington, TX 76019, USA}

\author[a]{A.~Ashtari Esfahani}
\author[b]{S.~B\"oser}
\author[a]{N.~Buzinsky}
\author[c]{M.~C.~Carmona-Benitez}
\author[a]{C.~Claessens} 
\author[c]{L.~de~Viveiros}
\author[b]{M.~Fertl}
\author[d]{J.~A.~Formaggio}
\author[e]{B.~T.~Foust}
\author[e]{J.~K.~Gaison}
\author[e]{M.~Grando}
\author[a]{J.~Hartse}
\author[f]{K.~M.~Heeger}
\author[e]{X.~Huyan}
\author[e]{A.~M.~Jones}
\author[g]{B.~J.~P.~Jones}
\author[h]{K.~Kazkaz}
\author[e]{B.~H.~LaRoque}
\author[d]{M.~Li}
\author[b]{A.~Lindman}
\author[a]{A.~Marsteller}
\author[b]{C.~Matth\'{e}}
\author[i]{R.~Mohiuddin} 
\author[i]{B.~Monreal}
\author[b]{B.~Mucogllava}
\author[c]{R.~Mueller}
\author[g]{A.~Negi}
\author[f]{J.~A.~Nikkel} 
\author[a]{E.~Novitski} 
\author[e]{N.~S.~Oblath}
\author[j]{M.~Oueslati}
\author[d]{J.~I.~Pe\~{n}a}
\author[j]{W.~Pettus} 
\author[b]{R.~Reimann}
\author[a]{R.~G.~H.~Robertson}
\author[a]{G.~Rybka}
\author[f]{L.~Salda\~{n}a}
\author[f]{P.~L.~Slocum}
\author[d]{J.~Stachurska}
\author[i]{Y.-H.~Sun}
\author[f]{P.~T.~Surukuchi}
\author[e]{J.~R.~Tedeschi}
\author[f]{A.~B.~Telles}
\author[b]{F.~Thomas} 
\author[b]{L.~A.~Thorne} 
\author[k]{T.~Th\"ummler} 
\author[d]{W.~Van~De~Pontseele}
\author[e,a]{B.~A.~VanDevender}
\author[f]{T.~E.~Weiss} 
\author[c]{T.~Wendler}
\author[c, 1]{A.~Ziegler, \note{Corresponding author.}}

\affiliation[a]{\uw}
\affiliation[b]{\mainz}
\affiliation[c]{\psu}
\affiliation[d]{\mitt}
\affiliation[e]{\pnnl}
\affiliation[f]{\yale}
\affiliation[g]{\uta}
\affiliation[h]{\llnl}
\affiliation[i]{\cwru}
\affiliation[j]{\iu}
\affiliation[k]{\kit}

%% file: sections/1-introduction.tex
\section{Introduction}
\label{sec:intro}
Cyclotron Radiation Emission Spectroscopy (CRES) is a technique for measuring the kinetic energies of charged particles by observing the frequency of the cyclotron radiation that is emitted as they travel through a magnetic field \cite{p8originalcres, p8prl2015}. The Project 8 Collaboration is developing the CRES technique as a next-generation approach to tritium beta-decay endpoint spectroscopy for neutrino mass measurement. Recently, Project 8 has performed the first CRES-based measurement of the tritium beta-decay energy spectrum and neutrino mass \cite{p8prl2023, p8prc2023}.

Previous CRES measurements have utilized relatively small volumes of radiation source gas that are directly integrated with a waveguide transmission line, which propagates the cyclotron radiation emitted by magnetically trapped electrons to a cryogenic amplifier. While this technology has had demonstrable success, it is not a feasible option for scaling up to larger measurement volumes. In particular, the goal of the Project 8 Collaboration is to use CRES combined with atomic tritium to measure the neutrino mass with a 40~meV sensitivity. Achieving this sensitivity goal will require a multi-cubic-meter scale measurement volume to obtain the required event statistics in the tritium beta-spectrum endpoint region; hence, there is a need for new techniques to enable large volume CRES measurements for future experiments.

One approach is to use antennas to collect a portion of the cyclotron radiation emitted by trapped electrons \cite{p8snowmass2022, p8jphysg}. A promising design is an inward-facing uniform circular array that surrounds a cylindrical containment volume. Increasing the size of the antenna array by adding additional rings of antennas along the longitudinal axis, allows one to grow the experiment volume until a sufficient amount of tritium gas can be observed by the array. A challenging aspect of this approach is that the total radiated power emitted by an electron near the tritium spectrum endpoint is on the order of 1~fW or less in a 1~T magnetic field. Because the CRES signal power and information is spread across the antenna array, detecting the presence of a CRES signal and determining the electron's kinetic energy requires reconstructing the entire array output over the duration of the event, posing a significant data acquisition and signal reconstruction challenge.

Previous measurements with the CRES technique have utilized a threshold on the frequency spectrum formed from a segment of time-series data. This algorithm relies on the detection of a frequency peak above the thermal noise background, which limits the kinematic parameter space of electrons available for reconstruction (see Section \ref{sec:bf-and-stft}). Although a trigger based on the amplitude of the frequency spectrum was adequate for previous Project 8 experiments, this approach does not provide sufficient detection efficiency for future CRES-based measurements of the neutrino mass. Better efficiency is possible by taking advantage of the deterministic CRES signal structure with a matched filter or machine learning based classifier \cite{p8ml_1}. In order to evaluate the relative gains in efficiency that come from utilizing these algorithms for antennas, analytical models that describe the detection performance of a power threshold and matched filter classifier are developed. In addition, a basic convolutional neural network (CNN) is implemented and tested as a first step towards the development of neural-network based classifiers for antenna array based CRES measurements. These results allow for a comparison between the estimated detection efficiencies of each of these methods, which are weighed against the associated computational costs for real-time applications.

The outline of this paper is as follows. Section \ref{sec:real-time-triggering} is an overview of a prototype antenna array CRES experiment, and describes the approach to real-time signal identification. Section \ref{sec:classifiers} develops models for the power threshold and matched filter algorithms and introduces the machine learning approach and CNN architecture. Section \ref{sec:method} describes the process for generating simulated CRES signal data and the details of training the CNN. Finally, Section \ref{sec:results} compares the signal classification accuracy for the three approaches and discusses the relevant trade-offs in terms of detection efficiency and computational cost.

\begin{figure}[h]
    \centering
    \begin{subfigure}{0.48\textwidth}
        \centering
        \includegraphics[width=0.9\textwidth]{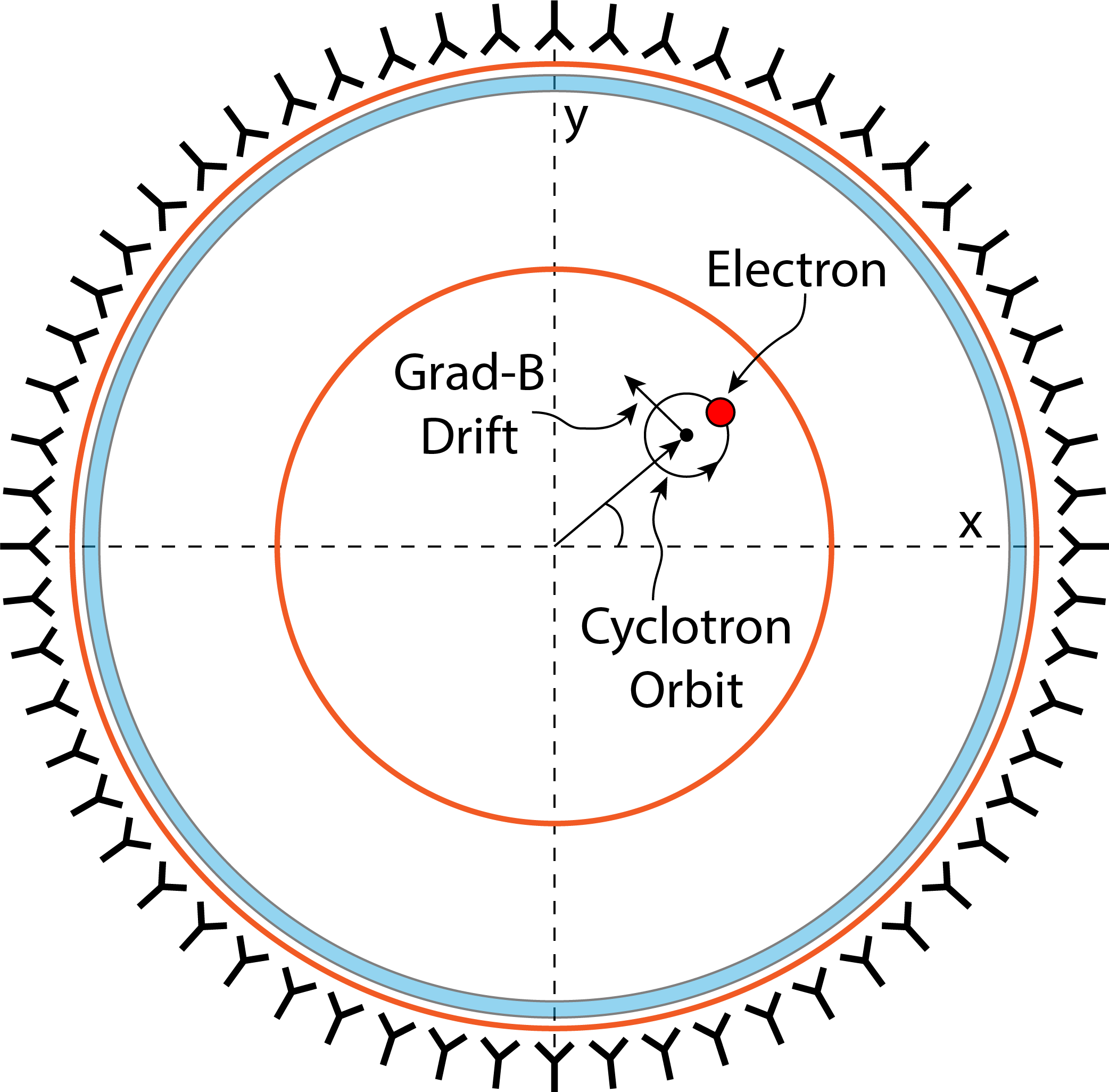}
        \caption{Top-down view.}
        \label{fig:apparatus_concept_top}
    \end{subfigure}
    \begin{subfigure}{0.48\textwidth}
        \centering
        \includegraphics[width=0.9\textwidth]{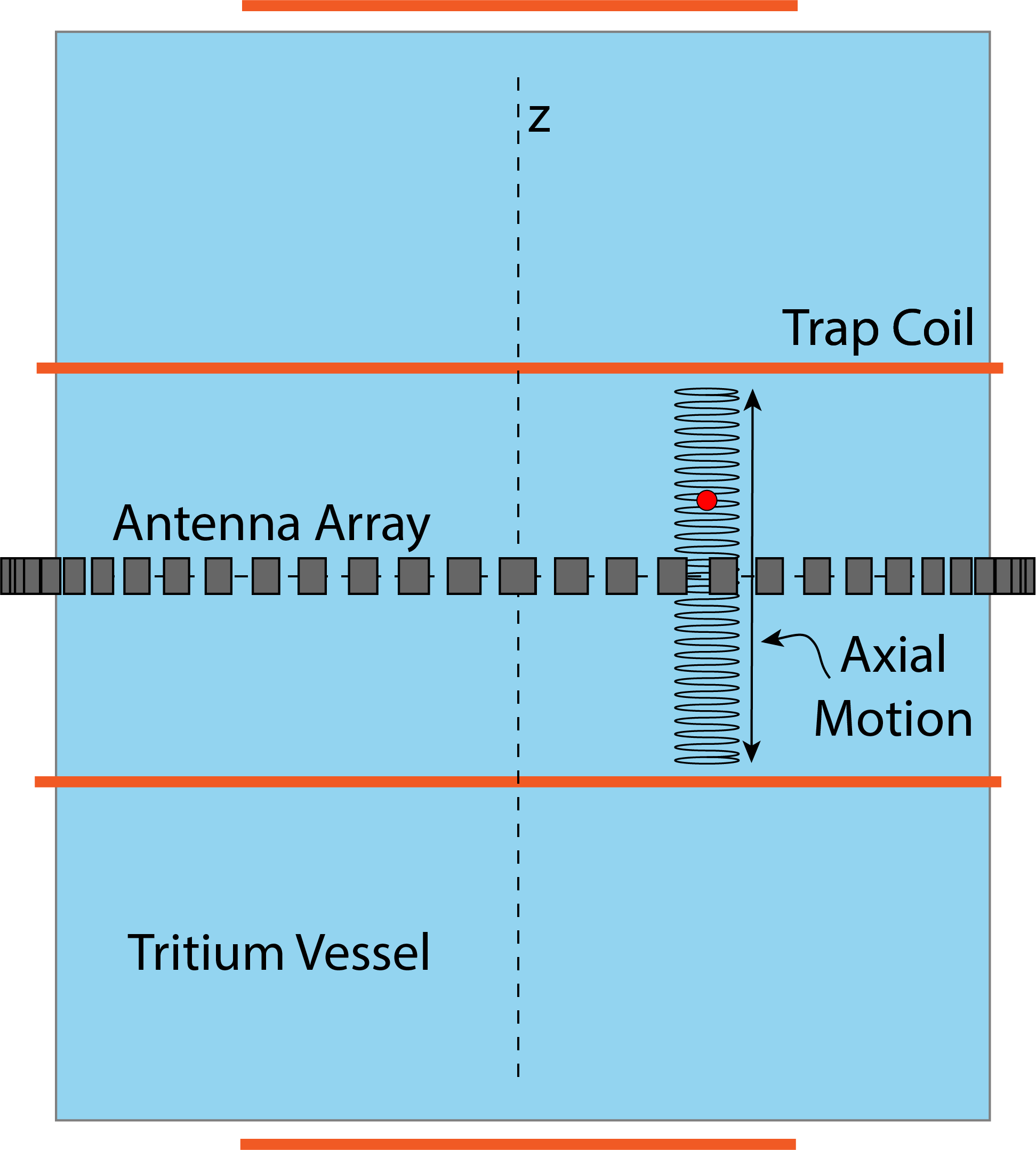}
        \caption{Side view.}
        \label{fig:apparatus_concept_side}
    \end{subfigure}
    \caption{An illustration of the conceptual design for the FSCD. The antenna array geometry consists of a 20~cm interior diameter with 60 independent antenna channels arranged evenly around the circumference. The nominal antenna design is sensitive to radiation in the frequency range of 25-26~GHz, which corresponds to the cyclotron frequency of electrons emitted near the tritium beta-spectrum endpoint in a 0.96~T magnetic field. The array is located at the center of the magnetic trap produced by a set of current-carrying coils.}
    \label{fig:apparatus_concept}
\end{figure}

%% file: sections/2-a_realtime_detection_routine.tex
\section{Signal Detection with Antenna Array CRES}
\label{sec:real-time-triggering}

\subsection{Antenna Array and Data Rate Estimates}
\label{sec:aa-and-daq}

In order to explore the potential of antenna array CRES for neutrino mass measurement, the Project 8 Collaboration has developed a conceptual design for a prototype antenna array CRES experiment \cite{p8PanicProc,p8snowmass2022}, called the Free-space CRES Demonstrator or FSCD (see Figure \ref{fig:apparatus_concept}). The FSCD design consists of a single ring of antennas, which is the simplest form of a uniform circular array configuration, to surround a radio-frequency (RF) transparent tritium gas vessel. A stand-alone version of this antenna array has been built and tested by the Project 8 collaboration \cite{p8jugaad} to validate simulations of the array radiation pattern and beamforming algorithms \cite{balanis}. In the FSCD the antenna array is positioned at the center of the magnetic trap formed by a set of electromagnetic coils, which create a local minimum in the magnetic field with a flat central region and steep walls in the radial and axial directions. The multi-coil trap design produces a trap with a larger volume than a two-coil bathtub trap or the single-coil traps used in the Project 8 Phase II experiment \cite{p8prl2023}.

When an electron is trapped its motion consists of three primary components. The component with the highest frequency is the cyclotron orbit whose frequency is determined by the size of the background magnetic field. The FSCD design assumes a background magnetic field value of approximately 0.96~T, which results in cyclotron frequencies of approximately 26~GHz for electrons with kinetic energies near the tritium beta-spectrum endpoint. The component with the next highest frequency is the axial oscillation experienced by electrons with pitch angles\footnote{Pitch angle is defined as the angle of the particle's total momentum with respect to the local magnetic field.} of less than $90^\circ$ as they move back and forth between the trap walls \cite{p8pheno}. Typical oscillation frequencies are on the order of $\sim10$ MHz, which results in an oscillation period that is a factor of $\sim10^2$ smaller than the observation time needed for precise measurement of the cyclotron frequency. Therefore, the axial extent of the electron's motion is unknown, and the electron is treated as if it is located in the average axial position at the bottom of the magnetic trap. The component of motion with the smallest frequency is the grad-B drift caused by radial field gradients in the trap, producing an orbit of the electron around the central axis of the trap with a frequency on the order of a few kHz, dependent on the pitch angle and the radial position of the electron. 

Each component of motion influences the shape of the cyclotron radiation signals received by the antenna array, therefore, the data acquisition (DAQ) system must be properly designed in order to resolve the effects of the cyclotron, pitch angle, and grad-B motions on the signal shape. Frequency down-conversion allows for sampling of the CRES signals with a bandwidth of 200~MHz, which must be large enough to contain all sidebands produced by pitch angle modulation. The noise temperature for the FSCD can be estimated using RF link-budget analysis, which depends upon the physical temperature of the experiment as well as the noise temperature of the cyrogenic amplifiers. The analysis presented in this paper assumes an effective system noise temperature of $\approx 10$~K, which is achievable with cyrogenic temperatures and low-noise commercially-available HEMT amplifiers.

\begin{figure}[htbp]
    \centering
    \includegraphics[width=0.9\textwidth]{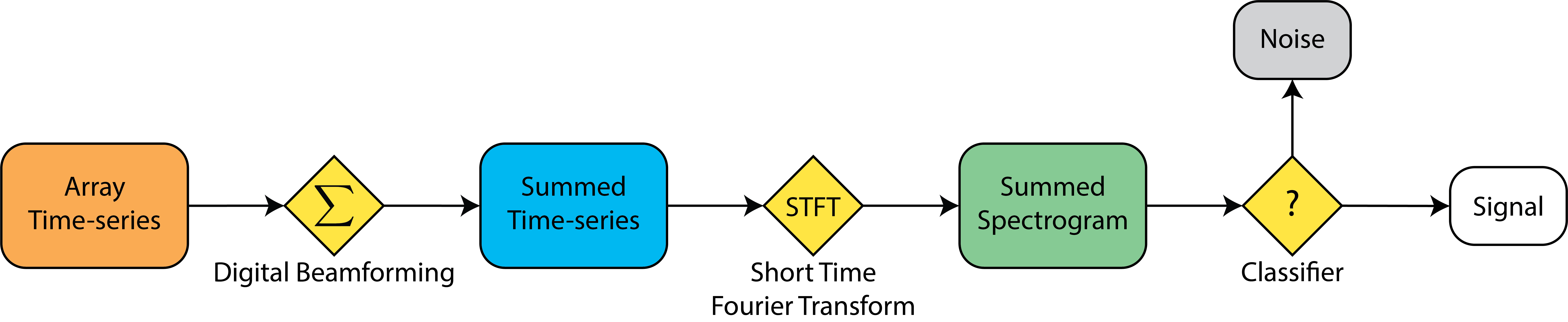}
    \caption{A block diagram illustration of the real-time signal detection algorithm proposed for antenna array CRES reconstruction.}
    \label{fig:signal_detection_routine}
\end{figure}

A design goal for the FSCD DAQ system is to enable a significant portion of the CRES event reconstruction to occur in real-time. The estimated data volume generated by the FSCD is 1~exabyte of raw data per year of operation, with the nominal array size of 60 antennas sampled at 200~MHz, which would be prohibitive to store for offline processing. Therefore, it is ideal to perform some CRES event reconstruction in real-time so that it is possible to save a reduced form of the data for offline analysis.

The first step of the real-time reconstruction would be a real-time signal detection algorithm, which is the focus of this paper. The basic approach consists of three operations performed on the time-series data blocks including digital beamforming, a short-time Fourier transform (STFT), and a binary classification algorithm to distinguish between data that consists of signal plus noise (Signal) and data that is purely noise (see Figure \ref{fig:signal_detection_routine}).

\subsection{Real-time Signal Detection}
\label{sec:bf-and-stft}

\begin{figure}[htbp]
    \centering
    \includegraphics[width=0.65\textwidth]{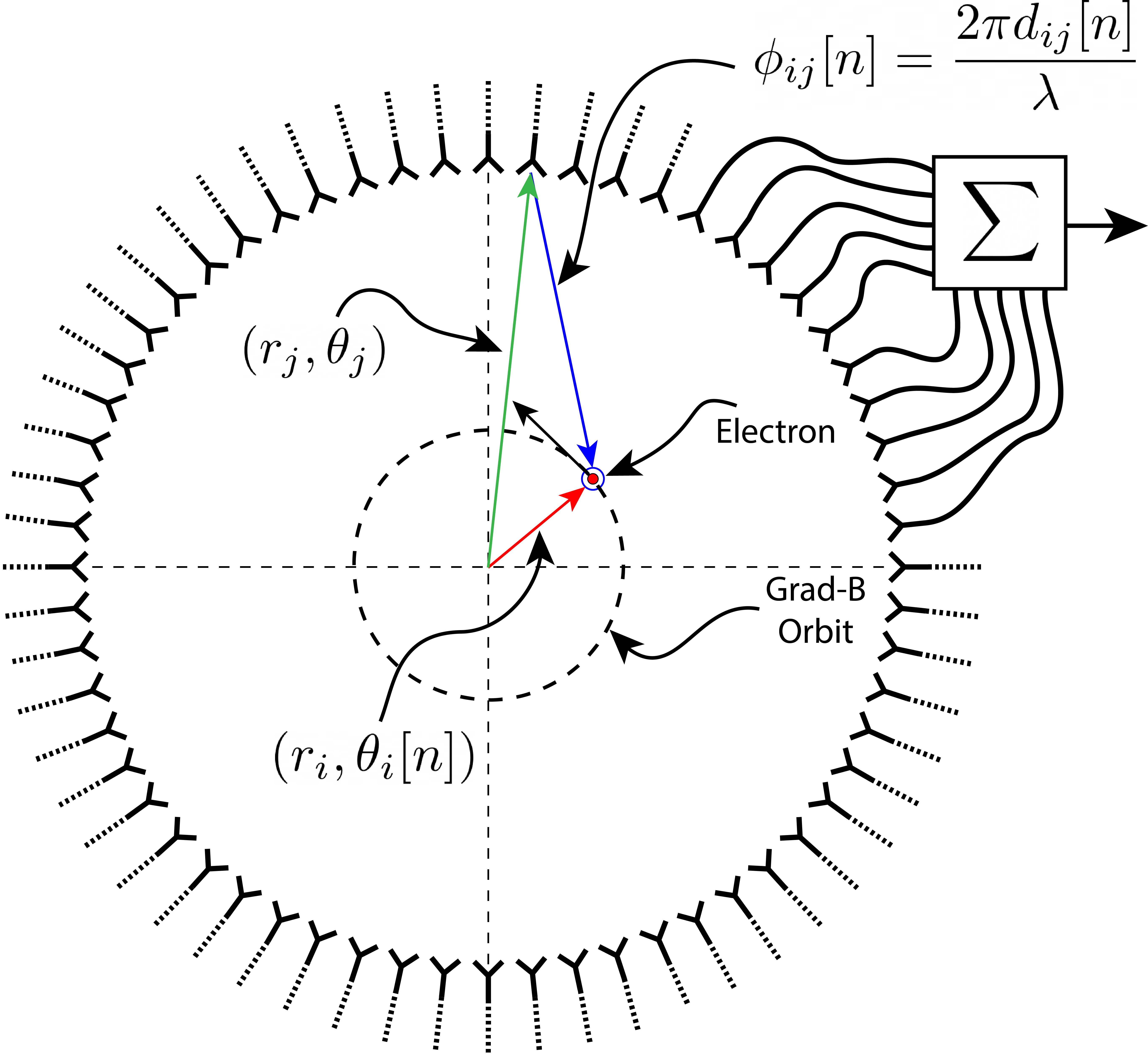}
    \caption{An illustration of the digital beamforming procedure. The blue arrow indicates the distance from the beamforming position to the antenna. In the configuration depicted the actual position of the electron matches the beamforming position, therefore, one expects constructive interference when the phase shifted signals are summed. To prevent the electron's grad-B motion from moving the electron off of the beamforming position, the beamforming phases include time-dependence to follow the trajectory of the electron in the magnetic trap.}
    \label{fig:beamforming}
\end{figure}

The first step in the real-time detection algorithm is digital beamforming, which is a phased summation of the signals received by the array (see Figure \ref{fig:beamforming}). The phase shifts correspond to the path length differences between a spatial beamforming position and each antenna such that, when there is an electron located at the beamforming position, all the signals received by the array constructively interfere. Since one does not know a priori where an electron will be produced in the detector, a grid of beamforming positions is designed to cover the entire azimuthal plane where electrons can be trapped. The phased summation is performed for all points in the grid at each time step. As discussed in Section \ref{sec:aa-and-daq}, the axial oscillation of the electrons prevents one from resolving its position along the z-axis, therefore, the beamforming grid need only cover the possible positions of the electron in the two-dimensional plane defined by the antenna array. 

Mathematically, digital beamforming can be expressed as
\begin{equation}
    y_i[n] = \sum_{j=1}^{N_\mathrm{ant}}\Phi_{ij}[n]x_j[n],
    \label{eq:beamforming}
\end{equation}
where $x_j[n]$ is the voltage sampled at antenna $j$ at time $n$, $\Phi_{ij}[n]$ is a matrix element from the time-dependent beamforming phase shift matrix, and $y_i[n]$ is the summed time-series for beamforming position $i$. The elements of the beamforming phase shift matrix can be expressed as a weighted complex exponential
\begin{equation}
    \Phi_{ij}[n]=A_{ij}[n]\exp{\left(i\phi_{ij}[n]\right)},
\end{equation}
where the weight $A_{ij}$ accounts for the relative power increase for antennas that are closer to the position of the electron, and $\phi_{ij}$ is the total beamforming phase shift for the $j$-th antenna and the $i$-th beamforming position. 

The beamforming phase shift is a sum of two terms
\begin{equation}
    \phi_{ij}[n]=\frac{2\pi d_{ij}[n]}{\lambda}+\theta_{ij}[n],
\end{equation}
where the first term is the phase shift originating from the path length difference ($d_{ij}[n]$) between the beamforming and antenna positions, which are represented by the vectors $(r_j,\theta_j)$ and $(r_i,\theta_i[n])$ (see Figure \ref{fig:beamforming}), and the second term is the angular separation ($\theta_{ij}[n]$) of the two positions. The angular separation enters into the beamforming phase due to an effect caused by the circular cyclotron orbit of the electron that produces radiation whose phase is linearly dependent on the relative azimuthal position of the antenna \cite{nb_thesis, p8synca}. The time-dependence of the beamforming phases corrects for the effects of the grad-B drift, which cause the guiding centers of electrons to orbit the center of the magnetic trap. The correction adds a linear time-dependence to the azimuthal beamforming position,
\begin{equation}
    \theta_{ij}[n]=\theta_j-\theta_i[n] = \theta_j - \omega_{\nabla B}t[n]+\theta_{i}[0],
\end{equation}
where $\theta_j$ is the fixed azimuthal position of antenna $j$, $\theta_{i}[0]$ is the starting azimuthal coordinate of the beamforming position, $t[n]$ is the time vector, and $\omega_{\nabla B}$ is the grad-B drift frequency, which allows the beamforming phases to track the XY-position of the guiding center. Predicting accurate values of $\omega_{\nabla B}$ for a specific trap and set of kinematic parameters can be done with simulations, which are performed using the Locust software package \cite{p8locustpaper} developed by Project 8.

\begin{figure}[ht]
    \centering
    \includegraphics[width=.7\textwidth]{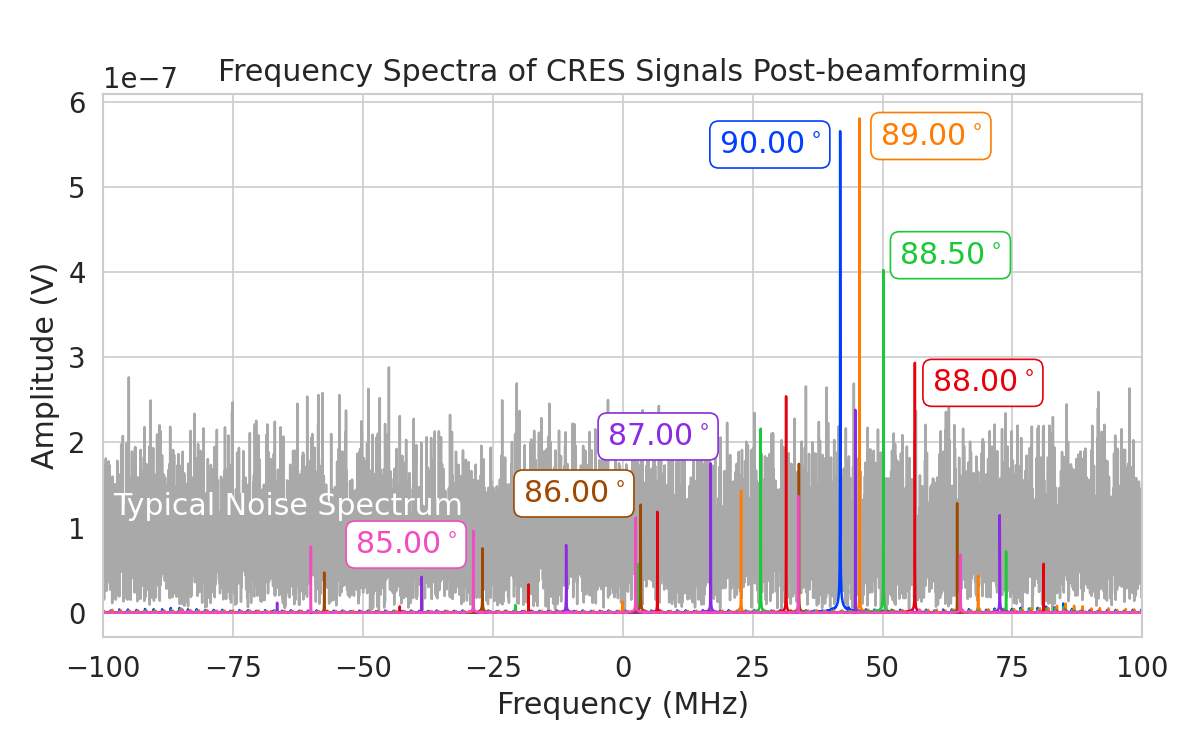}
    \caption{Frequency spectra of simulated CRES events in the FSCD magnetic trap after beamforming. The depicted frequency range has been down-converted using a local oscillator frequency of 25.86~GHz. The signal of a $90^\circ$ electron consists of a single frequency component that is clearly detectable using a power threshold on the frequency spectrum. This power threshold remains effective for signals with relatively large pitch angles such as $89.0^\circ$, which are composed of a main carrier and a few small sidebands. Signals with smaller pitch angles, below about $88.5^\circ$, are dominated by sidebands such that no single frequency component can be distinguished from the noise with sufficient statistical confidence using a power threshold.
    }
    \label{fig:signal_post_bf_example}
\end{figure}

After digital beamforming, a STFT is applied to the summed time-series to obtain the signal frequency spectrum (see Figure \ref{fig:signal_post_bf_example}). The sparseness of CRES signals in the frequency domain makes this representation better than the time domain for the purposes of signal detection. The frequency spectra of CRES signals are well-approximated by a frequency and amplitude modulated sinusoidal whose carrier frequency increases as a linear chirp \cite{p8pheno}. The modulation is caused by the axial oscillation of the electron in the magnetic trap, and the linear chirp is caused by the energy loss due to cyclotron radiation, which results in a relatively slow increase in the frequency components of the CRES signal over time. A typical CRES signal increases in frequency by approximately 15~kHz during the standard Fourier analysis window of 40.96~$\mu$sec, which is smaller than the frequency bin width for a 200~MHz sample rate. Therefore, when considering a single frequency spectrum it is justifiable to neglect the effects of the linear frequency chirp. 

The majority of the CRES signal power for electrons in the FSCD trap is contained in a single frequency component when the electron has a pitch angle $\gtrsim 88.5^\circ$. The remaining signal power is distributed between a small number of sidebands with amplitudes proportional to the electron's axial modulation (see Figure \ref{fig:signal_post_bf_example}). Signal detection for these pitch angles is straightforward using a simple power threshold on the STFT, since the amplitude of the main signal peak is well above the thermal noise spectrum. However, as the pitch angle of the electron is decreased below $88.5^\circ$, the maximum amplitude of the frequency spectrum becomes comparable to typical noise fluctuations. At this point, the power threshold trigger is no longer able to distinguish between signal and noise leading to a reduction in detection efficiency, which is directly linked to the neutrino mass sensitivity of the FSCD. Because the distribution of electron pitch angles is effectively uniform, utilizing a signal detection algorithm that can improve efficiency for pitch angles less than $88.5^\circ$ can lead to improvements in the neutrino mass sensitivity of the FSCD.

%% file: sections/3-signal-classifiers.tex
\section{Signal Detection Algorithms}
\label{sec:classifiers}

Modeling detection performance requires one to pose the signal detection problem in a consistent manner. The approach taken here uses the frequency spectra obtained from a STFT applied to beamformed time-series from the FSCD to perform a binary hypothesis test. Mathematically, this is expressed as,
\begin{align}
    \mathcal{H}_0 & : y[n]=\nu[n]\\
    \mathcal{H}_1 & : y[n]=x[n]+\nu[n].
\end{align}
Under hypothesis $\mathcal{H}_0$ the vector representing the frequency spectrum ($y[n]$) is composed only of complex white Gaussian noise (cWGN, $\nu[n]$) with total variance $\tau$, and under hypothesis $\mathcal{H}_1$ the frequency spectrum is composed of a CRES signal ($x[n]$) with added cWGN. The dominant noise source for the FSCD is expected to be thermal Nyquist-Johnson noise, which is well approximated by a cWGN distribution. The hypothesis test is performed by calculating the ratio between the log-likelihood probability distributions for the classifier under $\mathcal{H}_1$ and $\mathcal{H}_0$, which is the standard Neyman-Pearson approach to hypothesis testing \cite{detection_theory}. The output of the log-likelihood ratio test is called the test statistic, which is used to assign the data as belonging to the noise or signal classes using a decision threshold on the test statistic value. 

In practice, the decision threshold is selected by finding the value of the test statistic that guarantees a tolerable rate of false positives. Given this false positive rate (FPR), one attempts to find a classifier that maximizes the true positive rate (TPR), which is the probability of correctly identifying if a piece of data contains a signal. Because FSCD signal classifiers will be used to evaluate the spectra of $O(10^2)$ beamforming positions every 40.96~$\mu$sec, there is a requirement that the signal classifiers with FPR significantly smaller than 1\% to minimize the number of false positives that must be filtered out in later stages of the CRES signal reconstruction chain.

\subsection{Power Threshold}

The power threshold detection algorithm uses the maximum amplitude of the frequency spectrum as the detection test statistic. Consider the $\mathcal{H}_0$ hypothesis where the signal is pure cWGN. The performance of the power threshold can be modeled by first analyzing a single bin in the frequency spectrum. The probability that the amplitude of a frequency bin falls below the decision threshold is given by the Rayleigh cumulative distribution function (CDF),
\begin{equation}
    \mathrm{Ray}(|z|;\tau)=1-\exp{\left(-|z|^2/\tau\right)},
\end{equation}
where $|z|$ represents the value of the decision threshold on the spectrum amplitude, and $\tau$ is the cWGN variance (defined below, Equation \ref{eq:cwgn_var}). Because the noise samples are independent and identically distributed (IID), the probability that all bins in the frequency spectrum fall below the threshold is the joint CDF formed by the product of each individual frequency bin CDF,
\begin{equation}
    F_0(|z|;\tau, N_\mathrm{bin})=\mathrm{Ray}(|z|;\tau)^{N_\textrm{bin}}.
    \label{eq:fft_spectrum_cdf0}
\end{equation}
Finally, the PDF for the power threshold classifier can be obtained by differentiating Equation \ref{eq:fft_spectrum_cdf0}.

The noise of a beamformed frequency spectrum is a summation of all the noise samples from the array channels. The received Nyquist-Johnson noise power for a single antenna is given by $k_BT\Delta f$, where $k_B$ is Boltzmann's constant, $T$ is the system noise temperature, and $\Delta f$ is the sample rate. The beamformed noise variance is increased by a factor of $N_\textrm{ch}$, where $N_\textrm{ch}$ is the number of antennas, caused by the summation of incoherent noise samples, however, the noise variance per frequency bin is decreased by a factor equal to the number of samples in the STFT ($N_\textrm{FFT}$). The amplitude weights applied during beamforming also affect the noise variance in a position dependent way, since the analysis presented in this work focuses on only a single spatial position (see Section \ref{sec:method}), it was decided to weigh the signal in each channel equally, which results in the same noise variance for all beamforming positions. The final expression for the noise variance that describes a beamformed frequency spectrum is given by 
\begin{equation}
    \tau = k_BT\Delta fN_\textrm{ch}R/N_\textrm{FFT},
    \label{eq:cwgn_var}
\end{equation}
where the system impedance ($R$) has been used to convert from power to voltage-squared.

\begin{figure}[htbp]
    \centering
    \includegraphics[width=0.7\textwidth]{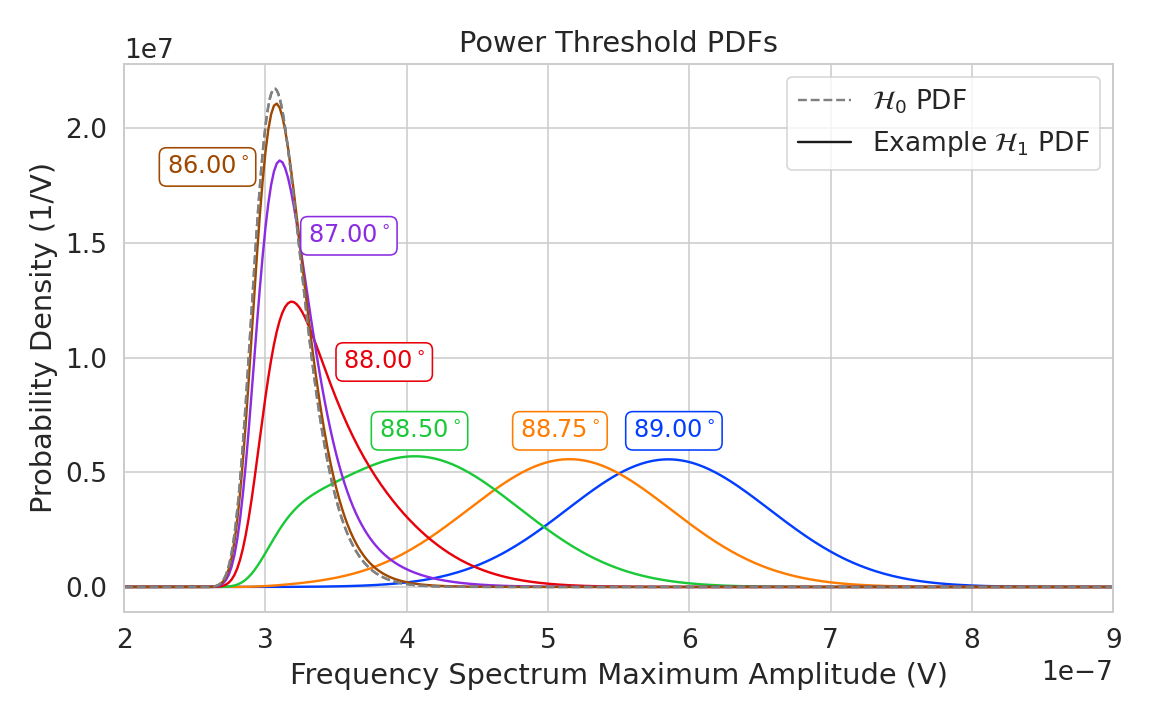}
    \caption{PDFs of the power threshold test statistic for CRES signals with various pitch angles as well as the PDF for the noise-only ($\mathcal{H}_0$) case. As the pitch angle is decreased the $\mathcal{H}_1$ PDFs converge towards the noise PDF, which indicates that the power threshold is unable to distinguish between signal and noise. }
    \label{fig:fft_pdf}
\end{figure}

Similar to $\mathcal{H}_0$, the power threshold classifier distribution under $\mathcal{H}_1$ is calculated by combining the distributions of individual frequency bins; however, the frequency bins that contain signal must be treated separately. The probability that the amplitude of a frequency bin which contains both signal and noise is below the decision threshold is described by a Rician CDF \cite{detection_theory}
\begin{equation}
    \mathrm{Rice}(|z|;\tau, |x_k|)=1-\int_{|z|}^{\infty}{d|\Tilde{z}|\frac{2|\Tilde{z}|}{\tau}\exp{\left(-\frac{|\Tilde{z}|^2+|x_k|^2}{\tau}\right)}\mathcal{I}_0\left(\frac{2|\Tilde{z}||x_k|}{\tau}\right)},
\end{equation}
where $|x_k|$ defines the amplitude of the $k$-th component of the signal frequency spectrum. The CDF that describes the probability that the entire spectrum falls below the decision threshold is the product of both signal and noise CDFs,
\begin{equation}
    F_1(|z|;\tau, \mathbf{x}_s, N_\mathrm{bin}, N_s)=\mathrm{Ray}(|z|;\tau)^{N_\mathrm{bin}-N_s}\prod_{k=1}^{N_s}{\mathrm{Rice}(|z|;\tau, \left|\mathbf{x}_s[k]\right|)}.
    \label{eq:fft_spectrum_cdf1}
\end{equation}
The first half of Equation \ref{eq:fft_spectrum_cdf1} is the contribution from the bins in the frequency spectrum that contain only noise, and the second half is the product of the Rician CDFs for the frequency bins that contain signal peaks with noise-free amplitudes of $\mathbf{x}_s=\left[x_1,\ldots, x_{N_s}\right]$, where $N_s$ is the number of non-zero frequency peaks in the CRES signal spectrum. Figure \ref{fig:fft_pdf} shows plots of example PDFs under $\mathcal{H}_1$ and $\mathcal{H}_0$.

\subsection{Matched Filtering}
\label{sec:classifiers-mf}

The shape of a CRES signal between random scattering events with the background gas is completely determined by the initial conditions of the electron, which implies that it is possible to apply matched filtering as a signal detection algorithm. A matched filter uses the shape of the known signal, which is called a template, to filter the incoming data by computing the convolution between the signal and the data \cite{detection_theory}. The matched filter is the optimal detector, which means it achieves the maximum TPR for a particular FPR, under the assumption that the signal is perfectly known and the noise is Gaussian distributed. Since CRES signals have an unknown shape but are deterministic, the matched filter can be applied by using simulations to generate a large number of signal templates, called a "template bank", which spans the parameter space of possible signals. Then at detection time, the template bank is used to identify signals by performing the matched filter convolution for each template in an exhaustive search.

CRES signals are highly periodic in nature. In such cases, it is advantageous to utilize the convolution theorem to replace the matched filter convolution with an inner product in the frequency-domain. Using the convolution theorem, the matched filter test statistic is given by
\begin{equation}
    \mathcal{T}=\max_{m}\left|\sum_{n=0}^{N_\mathrm{bin}}h_m^\dagger[n]y[n]\right|,
    \label{eq:mf_test_stat}
\end{equation}
where $h_m^\dagger[n]$ is the complex conjugate of the $m$-th signal template and $y[n]$ is the frequency spectrum of the beamformed time-series. 

\subsubsection*{Single Template}

The approach to deriving PDFs that describe the matched filter template bank will be to first derive PDFs for $\mathcal{H}_0$ and $\mathcal{H}_1$ in the case of a single template and use these solutions to create PDFs that describe the multi-template case. In the case when the template bank consists of only a single template it is possible to derive an exact analytical form for the PDF. Consider the $\mathcal{H}_1$ case, where the equation describing the matched filter test statistic, also known as the matched filter score, becomes
\begin{equation}
    \mathcal{T}=\left|\sum_{n=0}^{N_\mathrm{bin}}h^\dagger[n]y[n]\right|.
    \label{eq:mf_inner_prod_1}
\end{equation}
Each noisy frequency bin is a sum of signal and cWGN, which means $y[n]$ is also a Gaussian distributed variable. Therefore, the value of the inner product between the template and the data is also a complex Gaussian variable; and, since the matched filter score is the magnitude of this inner product, it must follow a Rician distribution. 

In Appendix \ref{app:mf-pdf} the exact form of the matched filter score PDF is derived. The solution is
\begin{equation}
    P_1(w;\mathcal{T}_0) = 2w\exp{\left(-\left(w^2+\mathcal{T}_0^2 \right)\right)}I_0(2w\mathcal{T}_0),
    \label{eq:mf_pdf_1}
\end{equation}
where $I_0$ is the zeroth-order modified Bessel function of the first kind and $w$ is the value of the matched filter score decision threshold. The shape of the matched filter score distribution is controlled by the parameter $\mathcal{T}_0$, which is effectively the value of the matched filter score if the data contained no noise. Without noise, the data vector reduces to the signal, $\mathbf{x}$, in which case Equation \ref{eq:mf_inner_prod_1} becomes the magnitude of an inner product between two vectors. The magnitude of an inner product can be expressed in terms of the magnitudes of the vectors and a constant that describes the degree of orthogonality between them. Applying this to Equation \ref{eq:mf_inner_prod_1}, one obtains
\begin{equation}
    \mathcal{T}_0=\left|\mathbf{h}^\dagger\cdot\mathbf{x}\right| = \left|\mathbf{h}\right|\left|\mathbf{x}\right|\Gamma,
    \label{eq:ideal_mf_score}
\end{equation}
where $\Gamma$ is a number that ranges from 0 to 1 and describes the orthogonality between $\mathbf{h}$ and $\mathbf{x}$. $\Gamma$ effectively quantifies how well the template matches the unknown signal in the data.

The matched filter score PDF under $\mathcal{H}_0$ is readily obtained from Equation \ref{eq:mf_pdf_1} by setting the value of $\mathcal{T}_\mathrm{0}$ to zero, since the data contains no signal in the noise case. Doing this, one obtains a Rayleigh distribution,
\begin{equation}
    P_0(w) = 2w\exp{\left(-w^2\right)}.
    \label{eq:mf_pdf_0}
\end{equation}

\subsubsection*{Multi-template}

Equations \ref{eq:mf_pdf_1} and \ref{eq:mf_pdf_0} describe the behavior of the matched filter test statistic under $\mathcal{H}_0$ and $\mathcal{H}_1$ for a single template. However, defining a PDF that describes the matched filter test statistic in the case of multiple templates is in general a mathematically intractable problem, since there is no guarantee of orthogonality between matched filter templates. This leads to correlations between the matched filter scores of different templates, because only one sample of noise is used to compute the matched filter scores of the template bank. 

In order to proceed, it is assumed that the matched filter scores for all templates are IID variables, which allows one to ignore correlations between templates. The overall effect of this will be an underestimate of the performance of the matched filter by over-estimating the required number of templates and the magnitude of the statistical trials penalty. The magnitude of this underestimation, while it cannot be predicted in advance, can be quantified using Monte-Carlo tests of the matched filter templates and randomly generated test signals.

\begin{figure}[htbp]
    \centering
    \includegraphics[width=0.7\textwidth]{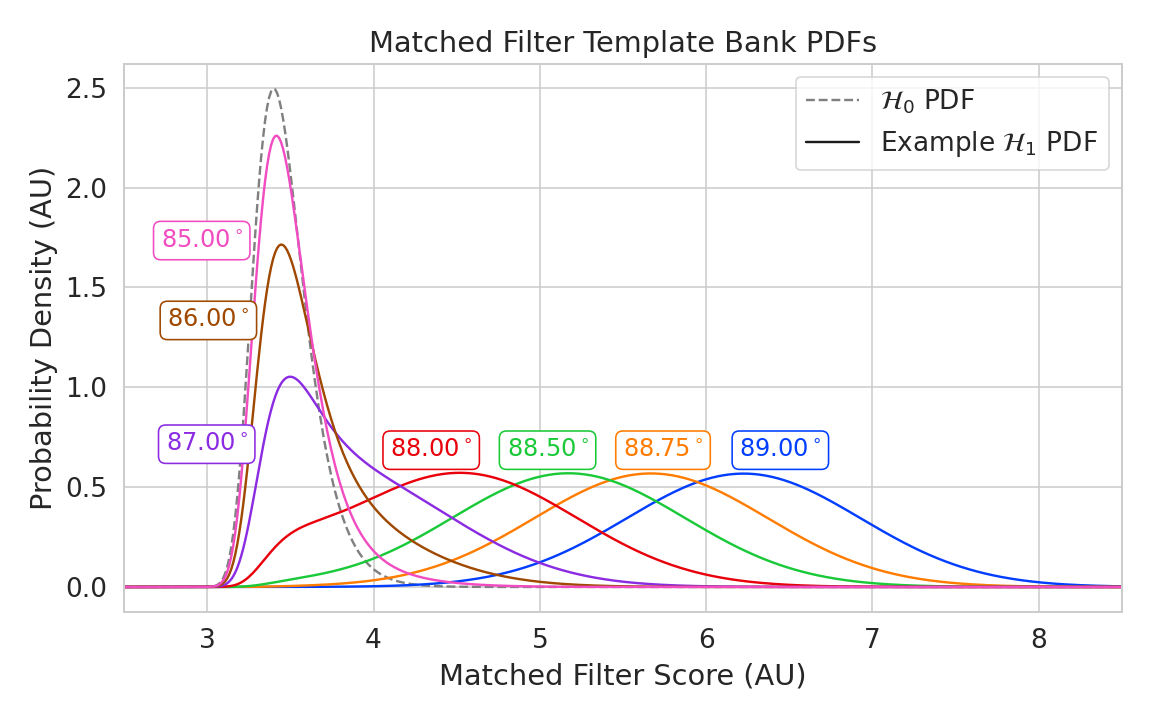}
    \caption{Example PDFs that describe the matched filter template bank test statistic for CRES signals with various pitch angles, as well as the estimated PDF for the noise-only case. $10^5$ matched filter templates and perfect match between signal and template, i.e. $\Gamma=1$, is assumed. One can observe that there is greater separation between the signal PDFs and the noise PDF for the matched filter template bank compared to the power threshold. Therefore, one expects that the matched filter will have generally better detection efficiency than the power threshold.}
    \label{fig:mf_pdf}
\end{figure}

The probability that the matched filter score falls below the decision threshold under $\mathcal{H}_0$ is again given by the CDF. Because of the assumption that matched filter scores from different templates are independent, the probability that the matched filter score for all templates falls below the threshold value is simply the joint CDF, which is
\begin{equation}
    F_{0}(w) = \left(1-e^{-w^2}\right)^{N_t},
    \label{eq:mf_joint_cdf_0}
\end{equation}
where $w$ is the matched filter score threshold and $N_t$ is the number of templates. One should expect that the distribution describing the maximum score of the matched filter template bank depends on $N_t$, because with more templates there is a greater chance of a random match between the template and noise data.

The CDF that describes $\mathcal{H}_1$ is derived by starting with the CDF of the best matching template. The best matching template is simply the template that yields the largest matched filter score. The score of the best template is defined by
\begin{equation}
    \mathcal{T}_\mathrm{best}=\max_{m}\left(|\mathbf{h}_m||\mathbf{x}|\Gamma_{m}\right) = |\mathbf{h}_\mathrm{best}||\mathbf{x}|\Gamma_{\mathrm{best}}.
\end{equation}
where variables $\mathbf{h}_\mathrm{best}$ and $\Gamma_\mathrm{best}$ are the template and corresponding match of the best-fitting template. A key performance parameter for a matched filter template bank is the mean value of $\Gamma_\mathrm{best}$ over the parameter space covered by the template bank. A higher density of matched filter templates will result in a higher $\overline{\Gamma_\mathrm{best}}$ and lead to better detection efficiency at the cost of a larger template bank.

The final form of the CDF under $\mathcal{H}_1$ is the joint distribution between the best matching template CDF and the CDFs of all other templates. By the orthogonality assumption described above, the matched filter scores for all other templates are treated as negligible ($\mathcal{T}_\mathrm{0}\approx0$). Therefore, the CDF for the matched filter template bank under $\mathcal{H}_1$ is simply
\begin{equation}
    F_{1}(w;\mathcal{T}_\mathrm{best})=F_\mathrm{best}(w;\mathcal{T}_\mathrm{best})\left(1-e^{-w^2}\right)^{N_t}.
    \label{eq:cdf1_mf}
\end{equation}
Figure \ref{fig:mf_pdf} shows plots of the matched filter template bank PDFs under $\mathcal{H}_0$ and $\mathcal{H}_1$.

\subsection{Machine Learning}
In this paper we specifically focus on the potential of Convolutional Neural Networks (CNN) as machine learning based signal classifiers at the trigger level. CNNs are constructed using a series of convolutional layers, each composed of a set of filters that are convolved with the input data. The individual convolutional filters can be viewed heuristically as matched filter templates \cite{cnn_are_mf} that are learned from a set of simulated data rather than being directly generated. This opens the possibility of finding a more efficient representation of the matched filter templates during the training process that can potentially reduce computational cost at inference time while retaining good classification performance. 

The machine learning approach is distinct from the power threshold and matched filtering in that there is no attempt to manually engineer a test statistic that can be computed from the input data. Instead, a test statistic is calculated by constructing a differentiable function that maps the complex frequency series to a binary classification as signal or noise. The differentiable function is trained using supervised learning to correctly perform this mapping. The test statistic for the machine learning classifier is expressed mathematically as
\begin{equation}
    \mathcal{T} = G(\mathbf{y};\mathbf{\Omega}),
\end{equation}
where $\mathbf{y}$ is the noisy data vector and $G(\mathbf{y}; \mathbf{\Omega})$ is the machine learning model parameterized by the weights $\mathbf{\Omega}$.

\begin{table}[h]
\centering
\caption{A summary of the CNN model layers and parameters. The output of each 1D-Convolution and Fully Connected layer is passed through a LeakyReLU activation function and re-normalized using batch normalization before being passed to the next layer in the model. The output of the final Fully Connected layer in the model is left without activation so that the model outputs can be directly passed to the Binary Cross-entropy loss function used during training. The first layer in the network has two input channels for the real and imaginary components of the spectrum. \label{tab:cnn_model_params}}
\smallskip
\begin{tabular}{@{}lllll@{}}
\hline
Layer&Type&Input Channels&Output Channels&Parameters\\
\hline
1 & 1D-Convolution & 2 & 15 & ($N_{\textrm{kernel}}=4$, $N_{\textrm{stride}}=1$)\\
2 & Maximum Pooling & 15 & 15 & ($N_{\textrm{kernel}}=4$, $N_{\textrm{stride}}=4$) \\
3 & 1D-Convolution & 15 & 20 & ($N_{\textrm{kernel}}=4$, $N_{\textrm{stride}}=1$)\\
4 & Maximum Pooling & 20 & 20 & ($N_{\textrm{kernel}}=4$, $N_{\textrm{stride}}=4$) \\
5 & 1D-Convolution & 20 & 25 & ($N_{\textrm{kernel}}=4$, $N_{\textrm{stride}}=1$)\\
6 & Maximum Pooling & 25 & 25 & ($N_{\textrm{kernel}}=4$, $N_{\textrm{stride}}=4$) \\
7 & Fully Connected & 3200 & 512 & NA \\
8 & Fully Connected & 512 & 64 & NA \\
9 & Fully Connected & 64 & 2 & NA \\
\hline
\end{tabular}
\end{table}

The CNN architecture used for this work is summarized by Table \ref{tab:cnn_model_params}. No strategic hyper-parameter optimization approach was implemented beyond the manual testing of different CNN architecture variations, so this particular model is best viewed as a proof-of-concept rather than a rigorously optimized design. Numerous model variations were tested, some with significantly more layers and convolutions filters per layer, as well as others that were even smaller than the architecture in Table \ref{tab:cnn_model_params}. Ultimately, the model architecture choice was driven by the motivation to find the minimal model whose classification performance was still comparable to the larger CNN's tested, because of the importance of minimizing computational cost in real-time applications. It is possible that more sophisticated machine learning models could improve upon the classification results achieved here, but this investigation is left for future work.

%% file: sections/4-methods.tex
\section{Methods}
\label{sec:method}

\subsection{Data Generation}
\label{sec:datasets}
Simulated CRES signals were generated using the Locust simulations package \cite{p8locustpaper, nb_thesis}. Locust uses the separately developed Kassiopeia package \cite{kassiopeia} to calculate the magnetic fields produced by a user-specified set of current carrying coils along with any specified background magnetic fields, resulting in a magnetic trap. Next, Kassiopeia calculates the trajectory of an electron in this magnetic field starting from a set of user specified initial conditions. The Locust software then uses the electron trajectories from Kassiopeia to calculate the resulting electromagnetic fields using the Li\'{e}nard-Wiechert equations, and determines the voltages generated in the antenna array with the antenna transfer function. Locust then simulates the down-conversion, filtering, and digitization steps resulting in the simulated CRES signals for an electron.

The shape of the received CRES signal is determined by the initial kinematic parameters, including the starting position of the electron, the starting kinetic energy of the electron, and the pitch angle. The studies performed here are constrained to a single initial electron position located at $(x,y,z)=(5, 0, 0)$~mm. Two datasets are generated using this starting position by varying the initial kinetic energy and pitch angle. 

The first dataset consists of a two-dimensional square grid spanning an energy range from 18575-18580~eV with a spacing of 0.1~eV, and pitch angles from $85.5$-$88.5^\circ$ with a spacing of $0.001^\circ$, resulting in 153051 signals with a unique energy-pitch angle combination. This dataset is intended to represent a matched filter template bank. The upper range of pitch angles is limited because of the greater relative detection efficiency of the matched filter and neural network classifiers in this pitch angle range. 

The second dataset was generated by randomly sampling uniform probability distributions covering the same parameter space to produce approximately 50000 signals randomly parameterized in energy and pitch angle. This dataset provides the training and test data for the machine learning approach, and acts as a representative sample of signals to evaluate the performance of the matched filter template bank.

Each signal was simulated for a duration of $40.96$~$\mu$s or 8192 samples starting at time $t=0$~s. This duration represents a single frequency spectrum generated by the STFT. The FSCD antenna array has sixty channels, and the output of the Locust simulations are a matrix of array snapshots with a size given by the number of channels times the event length ($N_\textrm{ch}\times N_\textrm{sample}$). The raw data from Locust is first summed using digital beamforming and converted to frequency spectra using a Fourier transform. The beamforming procedure uses the exact position and grad-B drift correction to simplify the comparison between trigger algorithms. Many beamforming positions would be used in practice and potentially several estimates of a typical $\omega_{\nabla B}$ depending on the variation of the grad-B drift frequency with pitch angle.

\subsection{Ensemble Averaging of Distributions}
\label{sec:ensemble_average}

As described above (see Section \ref{sec:datasets}), the parameter region of interest spans a 2-dimensional grid in energy and pitch angle, and the quantity of interest is the classifier's overall detection efficiency across this range. Equations \ref{eq:cdf1_mf} and \ref{eq:fft_spectrum_cdf1} were derived under the assumption that a particular signal was present in the data. Therefore, in order to describe the overall efficiency of the classifiers, we perform an ensemble average of the distributions, given by Equations \ref{eq:cdf1_mf} and \ref{eq:fft_spectrum_cdf1}, that describe detection probabilities for individual electrons in the dataset. From this set of distributions, we can obtain a single distribution the describes the overall detection efficiency of the classifier by performing an ensemble average. This averaging is performed (see Section \ref{sec:results}) using the second dataset described in Section \ref{sec:datasets}, which is randomly parameterized in energy and pitch angle.

\subsection{Template Bank Mean Match Estimation}
\label{sec:match}

In section \ref{sec:classifiers-mf}, we introduced $\overline{\Gamma_\mathrm{best}}$ as a figure-of-merit for a matched filter template bank. Generally, as the number and density of the matched filter templates is increased towards infinity $\overline{\Gamma_\mathrm{best}}\rightarrow1$, since it becomes increasingly likely that the signal will match at least one template. In the opposite extreme, where the template bank contains only one template, then $\overline{\Gamma_\mathrm{best}}\approx0$, since it is unlikely that any particular template will match a random signal from the experiment. The size of a matched filter template bank will always be limited based on practical constraints such as the availability of computational resources; therefore, it's undesirable to use more templates than what is required to achieve a $\overline{\Gamma_\mathrm{best}}$ that is close to one. 

\begin{figure}[htbp]
    \centering
    \includegraphics[width=0.69\textwidth]{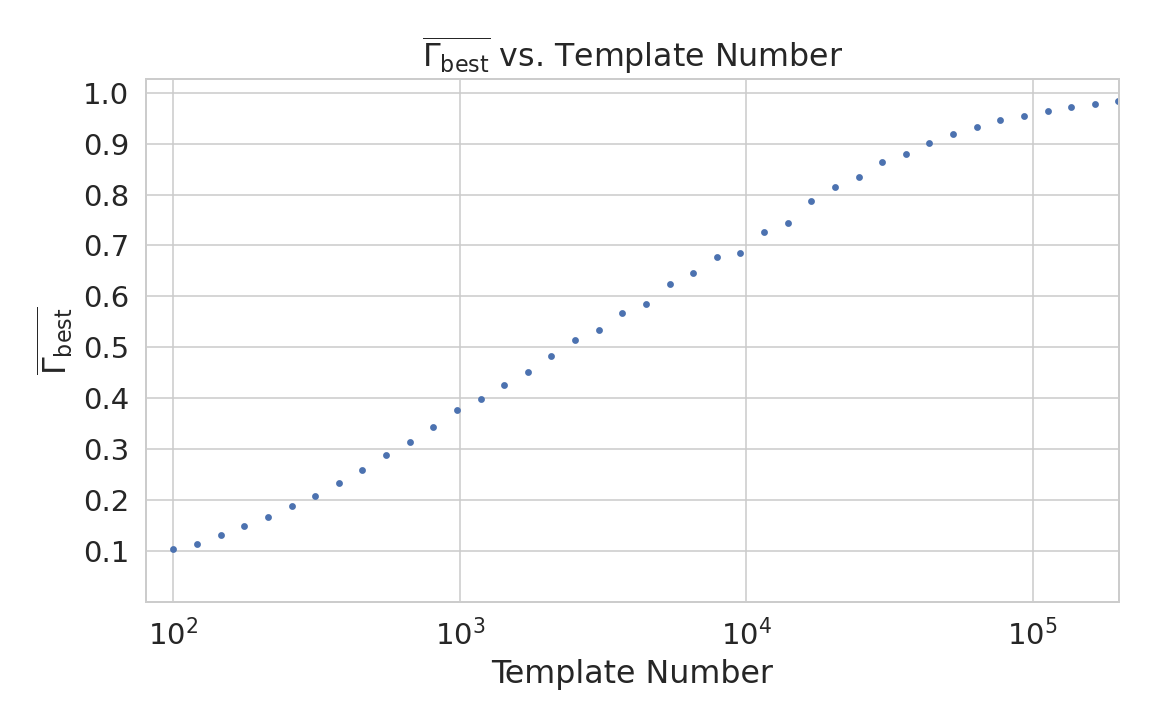}
    \caption{The mean match of the matched filter template bank to a test set of randomly parameterized signals as a function of the number of templates. The parameter space includes pitch angles from $85.5-88.5^\circ$ and energies from $18575-18580$~eV.}
    \label{fig:match_vs_template_number}
\end{figure}

In this work we elected to evaluate the detection efficiency of a template bank where $\overline{\Gamma_\mathrm{best}}=0.95$. The number of templates required to achieve this value of mean match was determined by Monte-Carlo studies using template banks of different sizes, which were obtained by decimating the regularly spaced dataset (see Section \ref{sec:datasets}) with series of integer factors. The randomly parameterized signals were used as test signals to calculate the mean match of the template bank by evaluating 
\begin{equation}
    \overline{\Gamma_\mathrm{best}}=\frac{1}{N_s}\sum_{l=1}^{N_s}{\frac{\mathcal{T}_\mathrm{best}}{\mathcal{T}_l}},
\end{equation}
where $N_s$ is the number of test signals and $\mathcal{T}_l$ is the ideal matched filter score for test signal $l$, which is obtained if there is perfect match between signal and template. 

The results of these calculations are summarized by Figure \ref{fig:match_vs_template_number}, which shows $\overline{\Gamma_\mathrm{best}}$ as a function of the number of templates ($N_t$). The linear trend of $\overline{\Gamma_\mathrm{best}}$ as a function of the logarithm of $N_t$ indicates that match scales exponentially with the size of the template bank, although this trend breaks down at large $N_t$ when match begins to saturate close to unity. Using linear interpolation between the data points, we identify $\approx10^5$ as the number of templates consistent with $\overline{\Gamma_\mathrm{best}}=0.95$.

\subsection{CNN Training and Data Augmentation}
The random dataset is split in half to create distinct training and test datasets for training the model. A randomly selected 20\% of the training data is isolated for use as a validation set during the training loop. The size of the training, validation, and test datasets are tripled by appending two additional copies of the data to increase the sample size of the dataset after data augmentation. A different sample of noise is added to the simulation data during the training loop, which prevents the model from overtraining on noise features. The training and test datasets contain an equal split between signal and noise data, which are randomly shuffled after each training epoch.

The Locust simulation data was augmented to make the datasets more representative of actual experiment data. As the signals are loaded for training a unique random phase shift is applied. Since the simulations are generated using the same initial axial position and cyclotron orbit phase, the randomization is an attempt to prevent overtraining on these features. During each training epoch the data is randomly shuffled and split into batches of 2500 signals. Each batch of signals is then circularly shifted by a random number of frequency bins to simulate a kinetic energy shift from $-75$ to $20$~eV, which imitates a dataset with a larger energy range. Next, a sample of cWGN, consistent with 10~K Nyquist-Johnson noise, is generated and added to the signal, which prevents overtraining on noise features. As a final step, the data is renormalized by the standard deviation of the noise so that the range of values in the data is close to $[-1,1]$, which ensures well-behaved back-propagation.

\begin{figure}[htbp]
    \centering
    \includegraphics[width=0.69\textwidth]{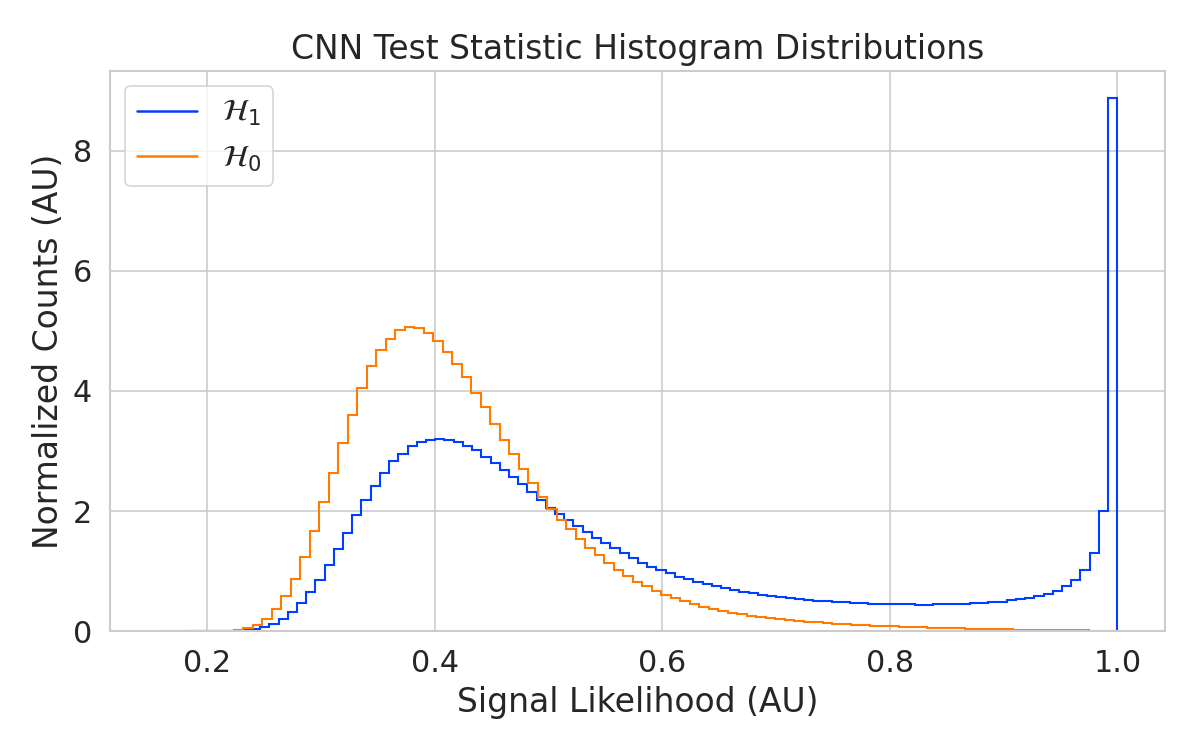}
    \caption{Histograms of the trained CNN model output from the test dataset. The blue histogram shows the model outputs for signal data. The oddly shaped peak near the end is the result of the softmax function mapping the long tail of the raw output distribution to the range $[0,1]$. }
    \label{fig:cnn_histogram}
\end{figure}

The Binary Cross-entropy loss function is used to compute the loss for each batch of data, and the model weights are updated using the ADAM optimizer with a learning rate of $5\times10^{-3}$. After each training epoch, the loss and classification accuracy of the validation dataset are computed to monitor for overtraining. It was noticed that because of the relatively high noise power and the fact that a new sample of noise was used for each batch, it was nearly impossible to over-train the model. Typically, the loss and classification accuracy of the model converged after a few hundred training epochs, but the training loop was extended to 3000 epochs to attempt to achieve the best possible performance. The training procedure generally took about 24~hrs using a single NVIDIA V100 Graphics Processing Unit (GPU) \cite{v100}.

After training the model, it was used classify the test dataset and generate histograms of the model outputs for both classes of data. The data augmentation procedure for the evaluation of the test data mirrors the training procedure without the validation split. Since a random circular shift and a new sample of WGN is added to each batch, the testing evaluation loop is run for 100 epochs to get a representative sample of noise and circular shifts. The model outputs are passed through a softmax activation and then combined into histograms (see Figure \ref{fig:cnn_histogram}).

%% file: sections/5-detection_performance_results.tex
\section{Results and Discussion}
\label{sec:results}

\subsection{Trigger Classification Performance}

\begin{figure}[htbp]
    \centering
    \includegraphics[width=0.7\textwidth]{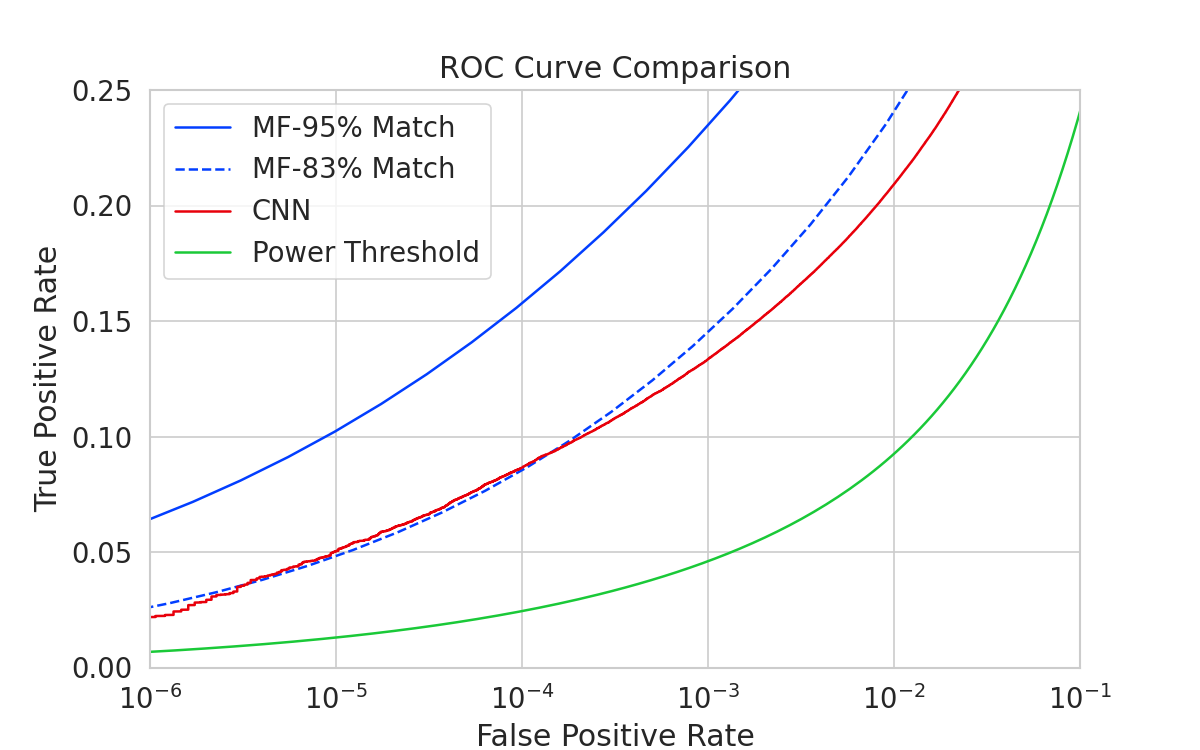}
    \caption{ROC curves describing the true positive rate (detection efficiency) for the three signal classification algorithms examined in this paper. The matched filter and power threshold curves are computed analytically using the distribution functions introduced in Section \ref{sec:classifiers}, and the CNN curve is computed numerically using the classification results on the test dataset. The percent match indicated in the legend refers to the value of $\overline{\Gamma_\mathrm{best}}$ for the template bank.
    }
    \label{fig:roc_compare}
\end{figure}
The detection performance of the signal classifiers can be compared by computing the receiver operating characteristic (ROC) curves (see Figure \ref{fig:roc_compare}).
A single ROC curve is obtained for the matched filter and power threshold classifiers by averaging over the distributions for individual CRES signals as described in Section \ref{sec:ensemble_average}. ROC curves are calculated for the matched filter using template banks of two different sizes corresponding to mean matches of 95\% and 83\%. The ROC curve describing the CNN is obtained numerically from the histograms of the model outputs for each signal class.

The true positive rates of the signal classifiers are equivalent to detection efficiency, and one sees that for the population of signals with pitch angles $<88.5^\circ$ the power threshold has a consistently lower detection efficiency than the CNN and the matched filter. This result might have been predicted from the visualization of signal spectra in Figure \ref{fig:signal_post_bf_example}, where it can be seen that a noise peak and a signal peak cannot be distinguished with high-confidence at small pitch angles. The CNN offers a significant and consistent increase in detection efficiency over the power threshold approach. 

If one compares the CNN to the matched filter, it can be seen that the performance of the tested network is roughly equivalent to a matched filter detector with a mean match of about 83\%, which uses approximately $2\times10^4$ matched filter templates. The overall best detection efficiency is achieved by the matched filter classifier if a large enough template bank is used. The plot displays the ROC curve for a matched filter template bank with 95\% mean match, which is achieved with approximately $10^5$ templates. Since the matched filter is known to be statistically optimal for detecting a known signal in WGN, it is unsurprising that this algorithm has the highest detection efficiency.

An important difference between the matched filter and CNN algorithms is that the CNN relies upon convolutions as its fundamental calculation mechanism, whereas our implementation of a matched filter utilizes an inner product. Since convolution is a translation invariant operation, the detection performance of CNN can be extended to a wider range of CRES event kinetic energies with less cost than the matched filter, a feature that is exploited during the CNN training by including circular translations of the CRES frequency spectra in the training loop. Increasing the range of detectable kinetic energies with a matched filter requires a proportional increase in the number of templates, which directly translates into increased computational and hardware costs. From a practical perspective, the detection algorithm is always limited by the available computational hardware, so estimating the relative costs is a key factor in determining their feasibility. A more detailed analysis of the relative costs of each of the detection algorithms is performed below.

\subsection{Computational Cost and Hardware Requirements}
\label{sec:dis-comp-cost}

The trade-off between better detection efficiency and computational cost is common to many signal detection problems and the FSCD is no exception. Computational costs can be related to actual hardware costs by calculating the theoretical amount of computer hardware required to implement the signal classifiers for real-time detection. The approach taken here utilizes order of magnitude estimates of the theoretical peak performance values for currently available GPUs as a metric. This approach underestimates the amount of required hardware, since it is unlikely that any CRES detection algorithm could reach the theoretical peak performance of the hardware. 

Since the signal detection algorithms are designed to work using beamformed frequency spectra, the computational cost of beamforming combined with a fast Fourier transform (FFT) is constant for all classifiers. The beamforming grid is assumed to contain $N_\mathrm{bf}$ beamforming positions, each of which will produce a frequency spectrum containing $N_\mathrm{bin}$ after the FFT. 

Considering the power threshold classifier, this results in $N_\mathrm{bin}N_\mathrm{b}$ frequency bins that must be checked every $N_\mathrm{bin}/f_\mathrm{s}$ seconds. The 20~cm diameter FSCD array requires $N_\mathrm{bf}\approx O(10^2)$ for sufficient coverage and has a sampling frequency $f_\mathrm{s}=200$~MHz with a Fourier analysis window of $N_\mathrm{bin}=8192$ samples. Therefore the power threshold requires approximately $O(10^{10})$~FLOPS to check in real-time with these parameters.

Current generations of GPUs have peak theoretical performances in the range of $O(10^{13})-O(10^{14})$~FLOPS \cite{h100}, dependent on the required floating-point precision of the computation. Therefore, the entire computational needs of a real-time triggering system using a power threshold classifier, including digital beamforming and generation of the STFT, could be met by a single high-end GPU or a small number of less powerful GPUs. Since triggering is only one step of the full real-time signal reconstruction approach, limiting the computational cost of this stage is ideal. However, the power threshold classifier does not provided sufficient detection efficiency across the entire range of possible signals, which is the primary motivation for exploring more complicated triggering solutions. 

As discussed, the computational cost of the matched filter approach requires counting the number of templates that must be checked for each frequency spectra produced by the STFT. Computing the matched filter scores requires $O(N_\mathrm{bf}N_\mathrm{t}N_\mathrm{bin})$ operations, since for each of the beamforming positions one must multiply $N_\mathrm{t}$ templates with a data vector that has length $N_\mathrm{bin}$. The computation must be performed in a time less-than or equal to $N_\mathrm{bin}/f_\mathrm{s}$ to keep up with the data generation rate. A 5~eV range of kinetic energies required $10^4$ to $10^5$ templates in order for the matched filter to exceed the performance of the CNN. The number of templates is expected to scale linearly with the total kinetic energy range of interest, therefore, $10^5$ to $10^6$ matched filter templates would be expected for the nominal 50~eV analysis window of the FSCD. Considering this, the estimated computational cost to implement a matched filter in a FSCD-scale experiment is between $O(10^{15})$ to $O(10^{16})$~FLOPS, which is $O(10^2)$ to $O(10^3)$ high-end GPUs .

The computational cost of the CNN can be estimated by simply summing the computational costs of the convolutions and matrix multiplications specified by the network architecture shown in Table \ref{tab:cnn_model_params}. Each convolutional layer consists of $N_\mathrm{in}N_\mathrm{out}N_\mathrm{kernel}L_\mathrm{input}$ floating-point operations, where $N_\mathrm{in}$ is the number of input channels, $N_\mathrm{out}$ is the number of output channels, $N_\mathrm{kernel}$ is the size of the convolutional kernel, and $L_\mathrm{input}$ is the length of the input vector, and the fully connected layers each contribute $N_\mathrm{in}N_\mathrm{out}$ operations. Summing all the neural network layers it is estimated that the CNN requires $O(10^6)$ floating point operations to evaluate each frequency spectra; therefore, the total computational cost of the CNN trigger is value multiplied by the number of beamforming positions per the data acquisition time, which is $O(10^{13})$~FLOPS or $O(10^0)$ GPUs.

Compared with the matched filter template bank approach the CNN requires $O(100)$ to $O(1000)$ fewer GPUs to implement, dependent on the exact number of templates used in the template bank. The 50~eV kinetic energy range is motivated by the application of these detection algorithms to an FSCD-like neutrino mass measurement experiment. However, if a significantly larger range of kinetic energies is required, a CNN may be the preferred detection approach despite the lower mean detection efficiency due to computational cost considerations.

Additional experiments with larger CNNs, generated by increasing the depth and width of the neural network, were performed. It was observed that these changes provided minimal ($\lesssim 1\%$) improvement in the classification accuracy of the model. A potential reason for this could be the sparse nature of the signals in the frequency domain and the low SNR, which makes for a challenging dataset to learn from. Future work might investigate modifications to the neural network architecture such as sparse convolutions, which may improve the classification accuracy of the model or further reduce the computational costs of this approach. Alternatively, more complicated CNN architectures such as a ResNet \cite{resnet, vgg} or VGG model may provide improved classification performance over a basic CNN. An additional promising area of investigation are recurrent neural networks, which may be able to exploit the time-ordered features of the STFT for more accurate signal detection if the electron signals last for multiple Fourier transform windows.

The estimate of the computational costs of the matched filter is somewhat naive if one notices that the majority of the values that make up a noise-free CRES frequency spectrum are zero (see Figure \ref{fig:signal_post_bf_example}). Therefore, the majority of operations in the matched filter inner product are unnecessary, and one could instead evaluate the matched filter inner product using only the $\lesssim10$ frequency peaks that make up the CRES signal. This optimization reduces the number of operations required to check each template by a factor of $O(100)$ to $O(1000)$, which brings the estimated computational cost of the matched filter in line with the CNN. Although this level of sparsity results in a multiplication with very low arithmetic complexity, the resulting sparse matched filter algorithm is still likely to be constrained by memory access speed rather than compute speed. Ultimately, the comparison of the relative computational and hardware costs between the matched filter and CNN will depend on the efficiency of the software implementation and hardware support for neural network and sparse matrix calculations, which will need to be determined using real-world benchmarks.

%% file: sections/6-conclusion.tex
\section{Conclusion}
\label{sec:conclusion}

Increasing the detection efficiency and overall event rate represents a key developmental path towards new scientific results and broader applications of the CRES technique. It is what motivates both the antenna array detection approach and the development of real-time signal reconstruction algorithms. The work presented here demonstrates that gains in detection efficiency are achievable by utilizing triggering algorithms that account for the specific shape of CRES signals in the detector. These algorithms emphasize the need for accurate and fast methods for CRES simulation, since they directly contribute to the success of matched filter methods by providing a way to generate expected signal templates and also serve as a source of training data for machine learning approaches. 

The down-side of more advanced approaches to signal detection and reconstruction is oftentimes the increase in computational resources required to implement them. However, it was shown that a CNN of minimal size was able to significantly improve detection performance above the baseline power threshold trigger algorithm with a theoretical computational cost of only $O(1)$ high-end GPU. This algorithm improves on detection performance while requiring at least a factor $O(10^2)$ less in computer relative to a matched filter template bank, which would be the classical approach to signal detection in Gaussian noise. Future work that obtains real-life benchmarks of the CNN and matched filter algorithms are required to support these conclusions, but this study has indicated that a real-time signal detection algorithm for an antenna array CRES experiment is computationally feasible without an extraordinary increase in resources.

It is worth emphasizing that, while this real-time signal detection algorithm has been developed specifically for the FSCD experiment, which uses a 60-channel array of antennas, this approach allows one to combine the signals from an array with an arbitrary number of receiving elements. Therefore, this general procedure could be implemented to perform signal reconstruction with close to optimal efficiency, especially if a matched filter classifier is used, for significantly larger antenna-based CRES experiments.

While this work has focused on the real-time detection of CRES signals from antenna arrays, these same signal classifiers could be used in CRES experiments utilizing different detector technologies, since the same principles of signal detection will apply. For example, previous CRES measurements by the Project 8 collaboration that utilized a waveguide gas cell, could in principle increase detection efficiency by employing a matched filter or neural network classifier to identify trapped electrons with pitch angles that are too small to be detected by the power threshold approach. Furthermore, alternative CRES detector technologies such as resonant cavities \cite{p8snowmass2022} could also see similar improvements in detection efficiency, which is of crucial importance to future efforts by the Project 8 collaboration to utilize CRES to measure the neutrino mass.

%% file: sections/appendix.tex
\appendix
\section{Derivation of the Matched Filter Score PDF}
\label{app:mf-pdf}

The matched filter template $\mathbf{h}$ is a simulated signal ($\mathbf{x}_h$) with a normalization factor
\begin{equation}
    \mathbf{h}=\frac{\mathbf{x}_h}{\sqrt{\tau|\mathbf{x}_h|^2}},
    \label{eq:appendix-mf-template}
\end{equation}
where $\tau$ is the noise variance. Inserting this into Equation \ref{eq:mf_test_stat} and expressing the data as a sum between a signal and a cWGN vector yields,
\begin{equation}
    \mathcal{T}=\frac{1}{\sqrt{\tau|\mathbf{x}_h|^2}}\left|\sum_{n=1}^{N_\mathrm{bin}}{x_h^\dagger[n]x[n]} + \sum_{n=1}^{N_\mathrm{bin}}{x_h^\dagger[n]\nu[n]}\right|.
    \label{eq:appendix-eqn-1}
\end{equation}

The first term is a scalar product between the signal and template vectors and the second term is a complex Gaussian distributed variable with variance one. For the purposes of identifying the statistical distribution, it is useful to rewrite the summation describing an inner product 
\begin{equation}
    \sum_{n=1}^{N_\mathrm{bin}}{x_h^\dagger[n]x[n]}=\mathbf{x}_h\cdot\mathbf{x}=|\mathbf{x}_h\cdot\mathbf{x}|e^{i\vartheta}\leq|\mathbf{x}_h||\mathbf{x}|e^{i\vartheta},
\end{equation}
the last step utilizes the Cauchy-Schwarz inequality, where equality is guaranteed when $\mathbf{x}=\mathbf{x}_h$. Instead of the inequality it is useful to define a quantity called "match" such that 
\begin{equation}
    |\mathbf{x}_h\cdot\mathbf{x}|e^{i\vartheta}=|\mathbf{x}_h||\mathbf{x}|\Gamma e^{i\vartheta},
\end{equation}
where the match factor $\Gamma\in[0,1]$. The match factor quantifies how well the template matches the signal.

The fact that the second term in Equation \ref{eq:appendix-eqn-1} is a random complex Gaussian variable with unity variance can be seen by noting that each of the noise samples are drawn from the complex Gaussian distribution, $\mathcal{N}(0,\tau)$. Therefore,
\begin{align}
    \frac{x_h^\dagger[n]}{\sqrt{\tau|\mathbf{x}_h|^2}}\nu[n]&\sim\mathcal{N}\left(0,\frac{x_h^\dagger[n]x_h[n]}{|\mathbf{x}_h|^2}\right),\\
    n=\sum_{n=1}^{N_\mathrm{bin}}{\frac{x_h[n]}{\sqrt{\tau|\mathbf{x}_h|^2}}\nu[n]}&\sim\mathcal{N}\left(0,\frac{\sum_{n=1}^{N_\mathrm{bin}}{x_h^\dagger[n]x_h[n]}}{|\mathbf{x}_h|^2}\right)=\mathcal{N}(0,1).
\end{align}

Equation \ref{eq:appendix-eqn-1} can now be simplified
\begin{equation}
    \mathcal{T}= \left||\mathbf{h}||\mathbf{x}|\Gamma e^{i\vartheta}+n\right|,
\end{equation}
where Equation \ref{eq:appendix-mf-template} has been used to redefine the inner product term. The quantity $|\mathbf{h}||\mathbf{x}|\Gamma$ is a real number, which is the matched filter score that one would expect if the data contained no noise. Since $\mathbf{h}$ and $\mathbf{x}$ can potentially be mismatched, $\Gamma$ is not necessarily equal to 1. The final simplification is to define $\mathcal{T}_\mathrm{0}=|\mathbf{h}||\mathbf{x}|\Gamma$, from which one obtains
\begin{equation}
    \mathcal{T}=|\mathcal{T}_\mathrm{0}e^{i\vartheta}+n|.
    \label{eq:appendix-simplified-mf-score}
\end{equation}

From Equation \ref{eq:appendix-simplified-mf-score} on can see that $\mathcal{T}$ is simply the magnitude of a complex number with added cWGN of variance 1, which follows the Rician distribution
\begin{equation}
    \mathrm{Rice}(x;\mathcal{T}_\mathrm{0}, 1/2)=2x\exp{\left(-\left(x^2+\mathcal{T}_\mathrm{0}^2\right)\right)}I_0\left(2x\mathcal{T}_\mathrm{0}\right).
\end{equation}

%% file: main.bbl
\providecommand{\href}[2]{#2}\begingroup\raggedright\begin{thebibliography}{10}

\bibitem{p8originalcres}
B.~Monreal and J.A.~Formaggio, \emph{Relativistic cyclotron radiation detection
  of tritium decay electrons as a new technique for measuring the neutrino
  mass}, \href{https://doi.org/10.1103/PhysRevD.80.051301}{\emph{Phys. Rev. D}
  {\bfseries 80} (2009) 051301}.

\bibitem{p8prl2015}
{\scshape Project 8} collaboration, \emph{{Single electron detection and
  spectroscopy via relativistic cyclotron radiation}},
  \href{https://doi.org/10.1103/PhysRevLett.114.162501}{\emph{Phys. Rev. Lett.}
  {\bfseries 114} (2015) 162501}
  [\href{https://arxiv.org/abs/1408.5362}{{\ttfamily 1408.5362}}].

\bibitem{p8prl2023}
{\scshape Project 8} collaboration, \emph{{Tritium Beta Spectrum Measurement
  and Neutrino Mass Limit from Cyclotron Radiation Emission Spectroscopy}},
  \href{https://doi.org/10.1103/PhysRevLett.131.102502}{\emph{Phys. Rev. Lett.}
  {\bfseries 131} (2023) 102502}
  [\href{https://arxiv.org/abs/2212.05048}{{\ttfamily 2212.05048}}].

\bibitem{p8prc2023}
{\scshape Project 8} collaboration, \emph{Cyclotron radiation emission
  spectroscopy of electrons from tritium beta decay and $^{83\rm m}$kr internal
  conversion},  \href{https://arxiv.org/abs/2303.12055}{{\ttfamily
  2303.12055}}.

\bibitem{p8snowmass2022}
{\scshape Project 8} collaboration, \emph{The project 8 neutrino mass
  experiment},  \href{https://arxiv.org/abs/2203.07349}{{\ttfamily
  2203.07349}}.

\bibitem{p8jphysg}
{\scshape Project 8} collaboration, \emph{{Determining the neutrino mass with
  cyclotron radiation emission spectroscopy\textemdash{}Project 8}},
  \href{https://doi.org/10.1088/1361-6471/aa5b4f}{\emph{J. Phys. G} {\bfseries
  44} (2017) 054004} [\href{https://arxiv.org/abs/1703.02037}{{\ttfamily
  1703.02037}}].

\bibitem{p8ml_1}
A.~Ashtari~Esfahani et~al., \emph{{Cyclotron radiation emission spectroscopy
  signal classification with machine learning in project 8}},
  \href{https://doi.org/10.1088/1367-2630/ab71bd}{\emph{New J. Phys.}
  {\bfseries 22} (2020) 033004}
  [\href{https://arxiv.org/abs/1909.08115}{{\ttfamily 1909.08115}}].

\bibitem{p8PanicProc}
{\scshape Project 8} collaboration, \emph{{Project 8: R\&D for a
  next-generation neutrino mass experiment}},
  \href{https://doi.org/10.22323/1.380.0283}{\emph{PoS} {\bfseries PANIC2021}
  (2022) 283}.

\bibitem{p8jugaad}
{\scshape Project 8} collaboration, \emph{Antenna arrays for physics
  measurements with large-scale cres detectors}, {\emph{In preparation} (2024)
  }.

\bibitem{balanis}
C.A.~Balanis, \emph{Antenna Theory: Analysis and Design, 4th Edition}, Wiley
  (2016).

\bibitem{p8pheno}
{\scshape The Project 8} collaboration, \emph{Electron radiated power in
  cyclotron radiation emission spectroscopy experiments},
  \href{https://doi.org/10.1103/PhysRevC.99.055501}{\emph{Phys. Rev. C}
  {\bfseries 99} (2019) 055501}.

\bibitem{nb_thesis}
N.~Buzinsky, \emph{Statistical Signal Processing and Detector Optimization in
  Project 8}, Ph.D. thesis, Massachusetts Institue of Technology2021.

\bibitem{p8synca}
A.A.~Esfahani, S.~Böser, N.~Buzinsky, M.~Carmona-Benitez, C.~Claessens,
  L.~de~Viveiros et~al., \emph{Synca: A synthetic cyclotron antenna for the
  project 8 collaboration},
  \href{https://doi.org/10.1088/1748-0221/18/01/P01034}{\emph{Journal of
  Instrumentation} {\bfseries 18} (2023) P01034}.

\bibitem{p8locustpaper}
{\scshape Project 8} collaboration, \emph{{Locust: C++ software for simulation
  of RF detection}}, \href{https://doi.org/10.1088/1367-2630/ab550d}{\emph{New
  J. Phys.} {\bfseries 21} (2019) 113051}
  [\href{https://arxiv.org/abs/1907.11124}{{\ttfamily 1907.11124}}].

\bibitem{detection_theory}
S.~Kay, \emph{Fundamentals of Statistical Signal Processing: Detection Theory,
  Volume II}, Pearson (1998).

\bibitem{cnn_are_mf}
L.~Stanković and D.~Mandic, \emph{Convolutional neural networks demystified: A
  matched filtering perspective-based tutorial},
  \href{https://doi.org/10.1109/TSMC.2022.3228597}{\emph{IEEE Transactions on
  Systems, Man, and Cybernetics: Systems} {\bfseries 53} (2023) 3614}.

\bibitem{kassiopeia}
D.~Furse et~al., \emph{{Kassiopeia: a modern, extensible C++ particle tracking
  package}}, \href{https://doi.org/10.1088/1367-2630/aa6950}{\emph{New Journal
  of Physics} {\bfseries 19} (2017) 053012}.

\bibitem{v100}
\url{https://www.nvidia.com/en-us/data-center/v100/}.

\bibitem{h100}
\url{https://www.nvidia.com/en-us/data-center/h100/}.

\bibitem{resnet}
K.~He, X.~Zhang, S.~Ren and J.~Sun, \emph{Deep residual learning for image
  recognition},  in \emph{2016 IEEE Conference on Computer Vision and Pattern
  Recognition (CVPR)}, pp.~770--7782016,
  \href{https://doi.org/10.1109/CVPR.2016.90}{DOI}.

\bibitem{vgg}
K.~Simonyan and A.~Zisserman, \emph{Very deep convolutional networks for
  large-scale image recognition},  in \emph{3rd International Conference on
  Learning Representations, {ICLR} 2015, San Diego, CA, USA, May 7-9, 2015,
  Conference Track Proceedings}, Y.~Bengio and Y.~LeCun, eds.2015,
  \href{http://arxiv.org/abs/1409.1556}{http://arxiv.org/abs/1409.1556}.

\end{thebibliography}\endgroup
